\documentclass[prl,twocolumn,showpacs,superscriptaddress,floatfix]{revtex4-1}
\usepackage{graphicx,amsfonts,amssymb,amsmath,xspace,dsfont}
\usepackage[colorlinks=true,citecolor=blue,linkcolor=red,urlcolor=blue]{hyperref}

\usepackage{color}
\definecolor{g}{rgb}{.1,0.4,.1} 
\definecolor{b}{rgb}{0,0.2,1}
\definecolor{rouge}{rgb}{0.82,0.,0.}
\definecolor{vert}{rgb}{0.,0.82,0.}
\definecolor{orange}{rgb}{1,0.5,0.}
\definecolor{bleu}{rgb}{0.,0.,0.82}
\definecolor{m}{rgb}{0.82,0.,0.82}
\definecolor{vert2}{rgb}{0.,0.5,0.}
\definecolor{rougeclair}{rgb}{1.0,0.7,0.7}

\begin{document}

\title{Topological Phase Transitions in the Golden String-Net Model}

\author{Marc Daniel Schulz}
\affiliation{Lehrstuhl f\"ur Theoretische Physik I, Technische Universit\"at Dortmund, Otto-Hahn-Stra\ss e 4, 44221 Dortmund, Germany}
\affiliation{Laboratoire de Physique Th\'eorique de la Mati\`ere Condens\'ee,
CNRS UMR 7600, Universit\'e Pierre et Marie Curie, 4 Place Jussieu, 75252
Paris Cedex 05, France}

\author{S\'{e}bastien Dusuel}
\affiliation{Lyc\'ee Saint-Louis, 44 Boulevard Saint-Michel, 75006 Paris, France}

\author{Kai Phillip Schmidt}
\affiliation{Lehrstuhl f\"ur Theoretische Physik I, Technische Universit\"at Dortmund, Otto-Hahn-Stra\ss e 4, 44221 Dortmund, Germany}

\author{Julien Vidal}
\affiliation{Laboratoire de Physique Th\'eorique de la Mati\`ere Condens\'ee,
CNRS UMR 7600, Universit\'e Pierre et Marie Curie, 4 Place Jussieu, 75252
Paris Cedex 05, France}

\begin{abstract}

We examine the zero-temperature phase diagram of the two-dimensional Levin-Wen string-net model
with Fibonacci anyons in the presence of competing interactions. Combining
high-order series expansions around three exactly solvable points and exact
diagonalizations, we find that the non-Abelian doubled Fibonacci topological phase is
separated from two nontopological phases by different second-order quantum critical
points, the positions of which are computed accurately. These trivial phases are
separated by a first-order transition occurring at a fourth exactly solvable
point where the ground-state manifold is infinitely many degenerate.
The evaluation of critical exponents suggests unusual universality classes.

\end{abstract}


\maketitle

%
%
%
%
Quantum phases of matter are often well described by local order parameters  and
Landau-Ginzburg symmetry-breaking theory is an efficient tool to analyze
transitions between these phases. However, in the late 1980s, a new class of
phases that cannot be understood in terms of local symmetries has emerged in the
context of high-temperature superconductivity \cite{Wen89_1,Wen89_2,Wen90_1}.
These phases, dubbed topological because of their sensitivity to the system
topology, have stimulated many studies in different domains (see
Ref.~\onlinecite{Wen12} for a recent review).  One of the most intriguing
properties of topologically ordered phases is that they are robust against
local (not too strong) perturbations \cite{Kitaev03,Bravyi10}. This stability makes
them especially appealing for quantum computation \cite{Preskill_HP} as well as
good candidates for quantum memories \cite{Dennis03}. Several experiments have
been proposed to realize the so-called topologically protected qubits \cite{Doucot12}.
In this perspective, a theoretical characterization of the robustness of
topological phases under strong perturbations as well as the nature of the
phase transitions signaling their breakdown is undoubtedly an important issue.
Thanks to recently proposed exactly solvable lattice models realizing various
topological phases of matter \cite{Kitaev03,Wen03,Levin05},  this program
has been undertaken for several models \cite{Trebst07,Hamma08,Kou08,Vidal09_1,Vidal09_2,Dusuel11,Wu12,Schulz12,Gils09_1,Gils09_3,Ludwig11,Poilblanc11,Freedman12,Burnell12}.

The main purpose of this Letter is to go one step beyond, by studying the phase diagram of a paradigmatic 2D non-Abelian model. We consider the Levin-Wen model \cite{Levin05} on the honeycomb lattice with Fibonacci anyons (the golden string-net model) in the presence of the same perturbation as the one introduced in Ref.~\onlinecite{Gils09_1}.  We determine the extension of the {\em doubled} Fibonacci (DFib) topological phase and show that it is separated from two other nontopological phases via second-order transitions that are analyzed in detail. 
%
%
\emph{Hilbert space ---}
%
%
Following the Levin-Wen construction, we consider a honeycomb lattice with anyonic degrees of freedom living on its edges. In the Fibonacci string-net model, these local (microscopic) degrees of freedom can be in two different states $|0\rangle$ or $|1\rangle$.
 The Hilbert space ${\mathcal H}$ is restricted to  states that satisfy the so-called branching rules stemming from the non-Abelian fusion rules
%
%
\begin{eqnarray}
0 \times a = a \times 0&=& a \:\: {\rm for} \:\: a \in \{0,1\}, \\
 1\times 1&=& 0+ 1. \label{eq:fusion}
\end{eqnarray} 
%
%
At each vertex of the honeycomb lattice, the fusion rules must not be violated; i.e., if one edge is in state $|1\rangle$, then at least one of the two other edges must be in the same state. For an arbitrary trivalent graph with $N_{\rm v}$  vertices, the dimension of the Hilbert space is then given by
%
%
\begin{equation}
\label{eq:dimH}
\dim \mathcal{H}= (1+\varphi^2)^{\frac{N_\mathrm{v}}{2}}+(1+\varphi^{-2})^{\frac{N_\mathrm{v}}{2}},
\end{equation} 
%
%
where $\varphi=\frac{1+\sqrt{5}}{2}$ is the golden ratio
(see, e.g., Ref.~\onlinecite{Simon12}).

%
%
\emph{Model.---}
%
%
We study the following Hamiltonian \cite{Gils09_1},
%
%
\begin{equation}
 \label{eq:ham}
H=- J_{\rm p} \sum_p  \delta_{\Phi(p),0} - J_{\rm e} \sum_e \delta_{l(e),0}.
\end{equation} 
%
%
The first term is  the string-net Hamiltonian introduced by Levin and
Wen \cite{Levin05}. It involves the projector $\delta_{\Phi(p),0}$ onto 
states with no flux $\Phi(p)$ through \mbox{plaquette $p$} \cite{Levin05, Gils09_1}.
The second term is diagonal in the basis introduced above since $\delta_{l(e),0}$
is the projector onto state $|0\rangle$ on edge $e$.

To help the reader to grasp the physical content of this Hamiltonian, let us
mention what happens if one replaces Eq.~(\ref{eq:fusion}) by the simpler
Abelian $\mathbb{Z}_2$ fusion rule $1\times1=0$. The model then becomes
Kitaev's toric code \cite{Kitaev03} on the honeycomb lattice, restricted to the
charge-free sector (because of the branching rules), in the presence of a
magnetic field $J_{\rm e}$ in the $x$ direction.  Indeed, one can write the Hamiltonian
in terms of Pauli matrices, with $\delta_{\Phi(p),0}=\big(\mathds{1}+\prod_{e\in
p} \sigma_e^z\big)/2$ and $\delta_{l(e),0}=\big(\mathds{1}+ \sigma_e^x\big)/2$.
Identifying plaquette fluxes with spin $1/2$ variables  as done in
Refs.~\onlinecite{Trebst07,Hamma08} for the square lattice, the Hamiltonian can
further be mapped onto the transverse-field Ising model on the triangular
lattice (of plaquettes), with coupling $J_{\rm e}$ and transverse-field $J_{\rm
p}$.

%
%
\emph{Limiting cases.---}
For convenience, let us set $J_{\rm p}=\cos \theta$ and $J_{\rm e}=\sin \theta$.
To our knowledge, contrary to the ladder geometry studied in
Ref.~\onlinecite{Gils09_1}, the Hamiltonian $H$ is exactly solvable only at the
four points for which $\theta$ is a multiple of $\pi/2$.
In the following, we discuss the low-energy spectrum of $H$ as well as the
corresponding phases around these special points and give some arguments in
favor of transitions between them. 

%
%
$(a)$ $\theta=0$~: For $J_{\rm p}>0$ and $J_{\rm e}=0$, the model reduces to the golden string-net model. Ground states $|\mathrm{g}\rangle$ are  flux-free states satisfying $\delta_{\Phi(p),0}|\mathrm{g}\rangle=|\mathrm{g}\rangle$ for all $p$ and thus have an energy per plaquette \mbox{$e_0=E_0/N_{\rm p}=-1$} ($N_{\rm p}$ being the number of plaquettes). Their degeneracy depends on the system topology which is the most salient property of a topological phase. 
For the Fibonacci string-net model on any trivalent graph, the ground state is unique on a sphere \cite{Hu12} whereas it is fourfold degenerate on a torus. Interestingly, one can also compute the degeneracy of the $k{\rm th}$ excited states (with energy  $E_k=E_0+k$)
%
%
\begin{equation}
\label{eq:deg_Fibo}
\mathcal{D}_k= 
\left(
\begin{array}{c}
N_{\mathrm p}
\\
k
\end{array}
\right) \left(p \: F_{k-1}^2+ q \:F_k^2+ r \:F_k F_{k-1}\right),
\end{equation} 
%
%
where we introduced the famous Fibonacci sequence defined for any integer $n$ by
$F_{n+1}=F_n+F_{n-1}$, with \mbox{$F_{-1}=1$} and $F_0=0$.
The integers $(p,q,r)$ depend on the surface considered. For instance, one has
\mbox{$(p,q,r)=(1,0,0)$} on a sphere whereas $(p,q,r)=(4,1,4)$ on a torus.
Equation~(\ref{eq:deg_Fibo}) shows that for the Fibonacci theory, {\em an odd number of
excitations can exist on a compact surface} contrary to the charge-free toric code discussed above
where fluxes are always created and annihilated by pairs. 
Note that the binomial coefficient simply arises from the
different ways to choose $k$ plaquettes carrying the flux excitations among
$N_{\mathrm p}$.

Products of {\em two} $F_k$'s stem from the fact that the
``emergent'' flux excitations are not the microscopic Fibonacci anyons
but are achiral combinations of {\em two} Fibonacci anyons (details will be given in Ref.~\onlinecite{Schulz13}).  The non-Abelian topological phase in the vicinity of $\theta=0$ is described by  a  DFib theory \cite{Fidkowski09,Wang_book,Levin05,Burnell10,Burnell11_1,Burnell11_2}. Excitations have a
trivial topological spin \cite{Levin05} and can fuse to the vacuum (also called trivial
particle) \cite{Koenig10}.  As such, they can also be considered
as bosons \cite{Bais09} and hence condense.

$(b)$ $\theta=\pi$~: For $J_{\rm e}=0$ and $J_{\rm p}<0$, the low-energy spectrum is very different. Indeed, in this case, the ground-state manifold is ${\mathcal D}_{N_{\rm p}}$-fold degenerate and spanned by all states $|g\rangle$ satisfying $\delta_{\Phi(p),0}|\mathrm{g}\rangle=0$ for all $p$. As discussed above, this degeneracy depends on the topology through its indices ($p,q,r$) so that one might be tempted to consider the system as topologically ordered. However, the local operator $\sum_e \delta_{l(e),0}$ couples the ground states and splits the degeneracy for any $J_{\rm e}\neq 0$. As a consequence, the system cannot be considered as topologically ordered \cite{Nussinov09}.

\begin{figure}[b]
\includegraphics[width= \columnwidth]{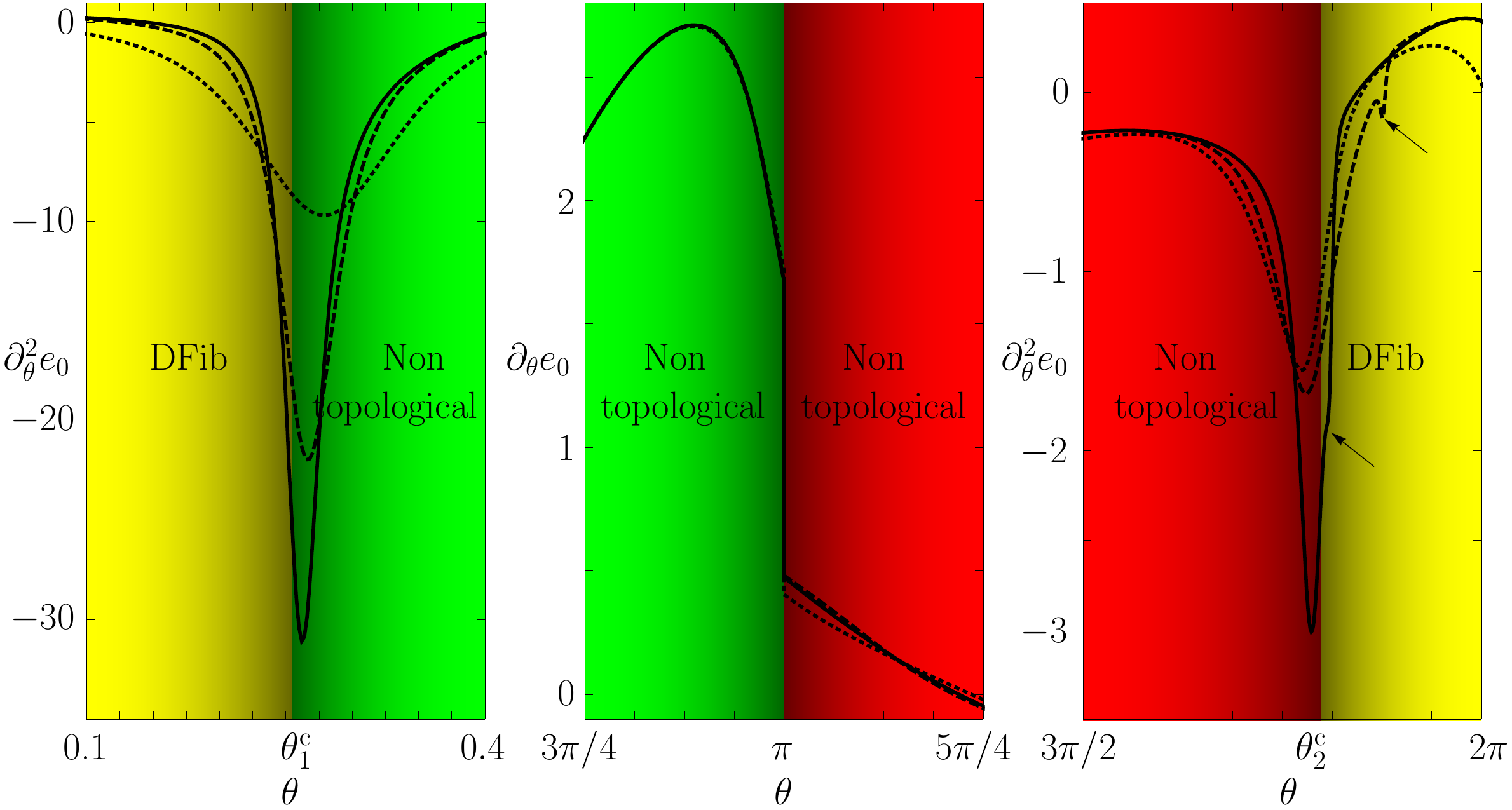}
\caption{(color online). ED results for $N_{\rm p}=2\times2$ (dotted line), $3\times 3$ (dashed line), and 
$\sqrt{13} \times \sqrt{13}$ (solid line) plaquettes.
Left and right panels~: $\partial_\theta^2 e_0$ decreases with the system size indicating second-order transitions at 
$\theta^{\rm c}_1$ and $\theta^{\rm c}_2$. Central panel~: $\partial_\theta e_0$ displays a clear jump at $\theta=\pi$ indicating a first-order transition. Dips indicated by arrows in the DFib topological phase are due to (irrelevant and avoided)  level crossings between the four lowest-energy levels that become degenerate in the thermodynamical limit on a torus.}
\label{fig:dgse}
\end{figure}

Owing to this huge degeneracy, we have not been able to analyze the vicinity of this point. However, numerical results obtained by exact diagonalizations clearly show that $(i)$ the degeneracy is lifted as soon as the coupling $J_{\rm e} \neq 0$ and $(ii)$ the ground state for $\theta=\pi^\pm$ is unique and adiabatically connected to the polarized ground states found at $\theta=\pi/2$ and $3\pi/2$, respectively, (see discussion below). This result is in stark contrast with the scenario described in Ref.~\onlinecite{Gils09_1} on the ladder where a gapless phase is observed for $\theta \in [\pi,3 \pi/2]$. 
In addition, as can be seen in Fig.~\ref{fig:dgse} (central panel), we found  a jump in $\partial_\theta e_0$ at $\theta=\pi$ (for all system sizes) indicating that the two gapped phases ($\theta=\pi^+$ and $\theta=\pi^-$) are separated by a first-order phase transition.

$(c)$ $\theta=\pi/2$~: For $J_{\rm p}=0$, the Hamiltonian $H$ is diagonal in the canonical basis of states satisfying the branching rules. For $J_{\rm e}>0$, the ground state is unique whatever the topology and corresponds to the fully polarized state where all edges are in the state $|0\rangle$ (with eigenenergy $E_0=-N_{\rm e}$, where $N_{\rm e}$ is the total number of edges).  First excited states are obtained by flipping six links around one hexagon. They behave as trivial hard-core bosons that become dynamical when the coupling $J_{\rm p}$ is switched on.
Thus, near $\theta=\pi/2$, the system is gapped but not topologically ordered, making the occurrence of a phase transition in the interval $[0,\pi/2]$ compulsory.

$(d)$ $\theta=3\pi/2$~: For  $J_{\rm e}<0$ and $J_{\rm p}=0$, the Hamiltonian is also diagonal  and the unique ground state is the fully polarized state where all edges are in the states $|1\rangle$ \mbox{($e_0=0$)}. Note that such a state  would be forbidden by the Abelian $\mathbb{Z}_2$ fusion rules.  First excited states are obtained from the ground state by flipping a single link. As previously, these localized excitations are trivial hard-core bosons that become mobile when $J_{\rm p}\neq 0$ so that one expects a phase transition in the interval $[3\pi/2,2\pi]$.

%
%
\begin{figure}[t]
\includegraphics[width= 0.9\columnwidth]{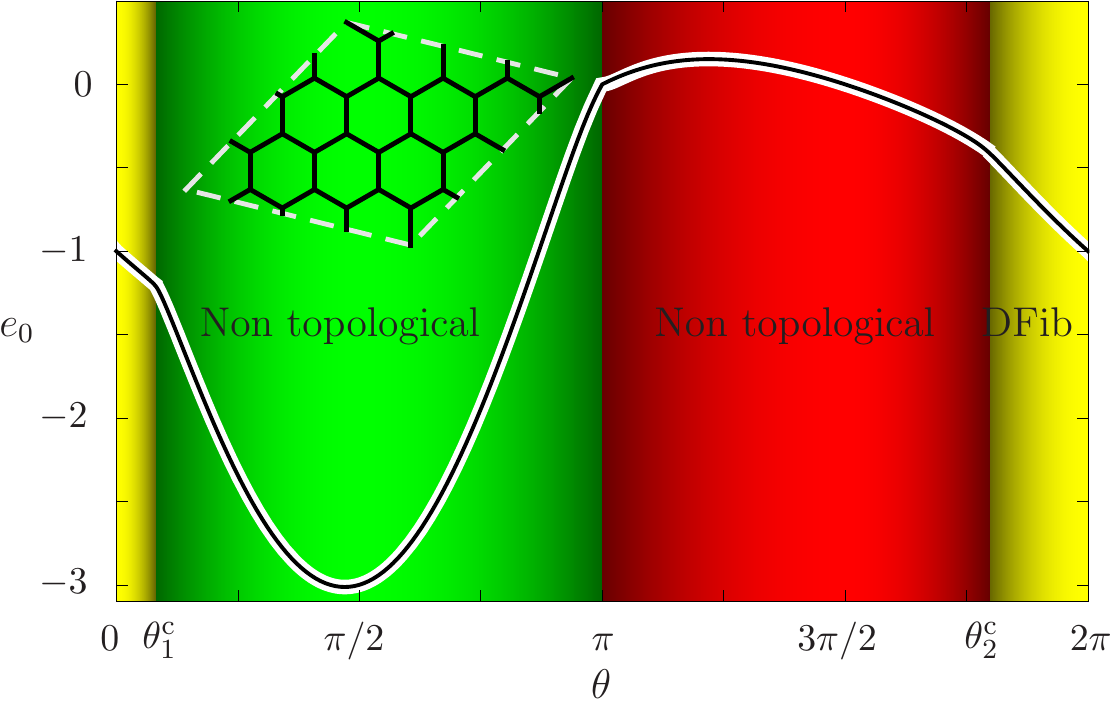}
\caption{(color online). 
Ground-state energy per plaquette $e_0=E_0/N_{\rm p}$ as a function of $\theta$. ED results (black line) for \mbox{$N_{\rm p}=\sqrt{13}\times\sqrt{13}$} plaquettes (see inset) are in excellent agreement with typical Pad\'e approximants (white lines) computed from high-order series expansions (see Supplemental Material).}
\label{fig:gse}
\end{figure}
%
%

\begin{figure}[t]
\includegraphics[width= \columnwidth]{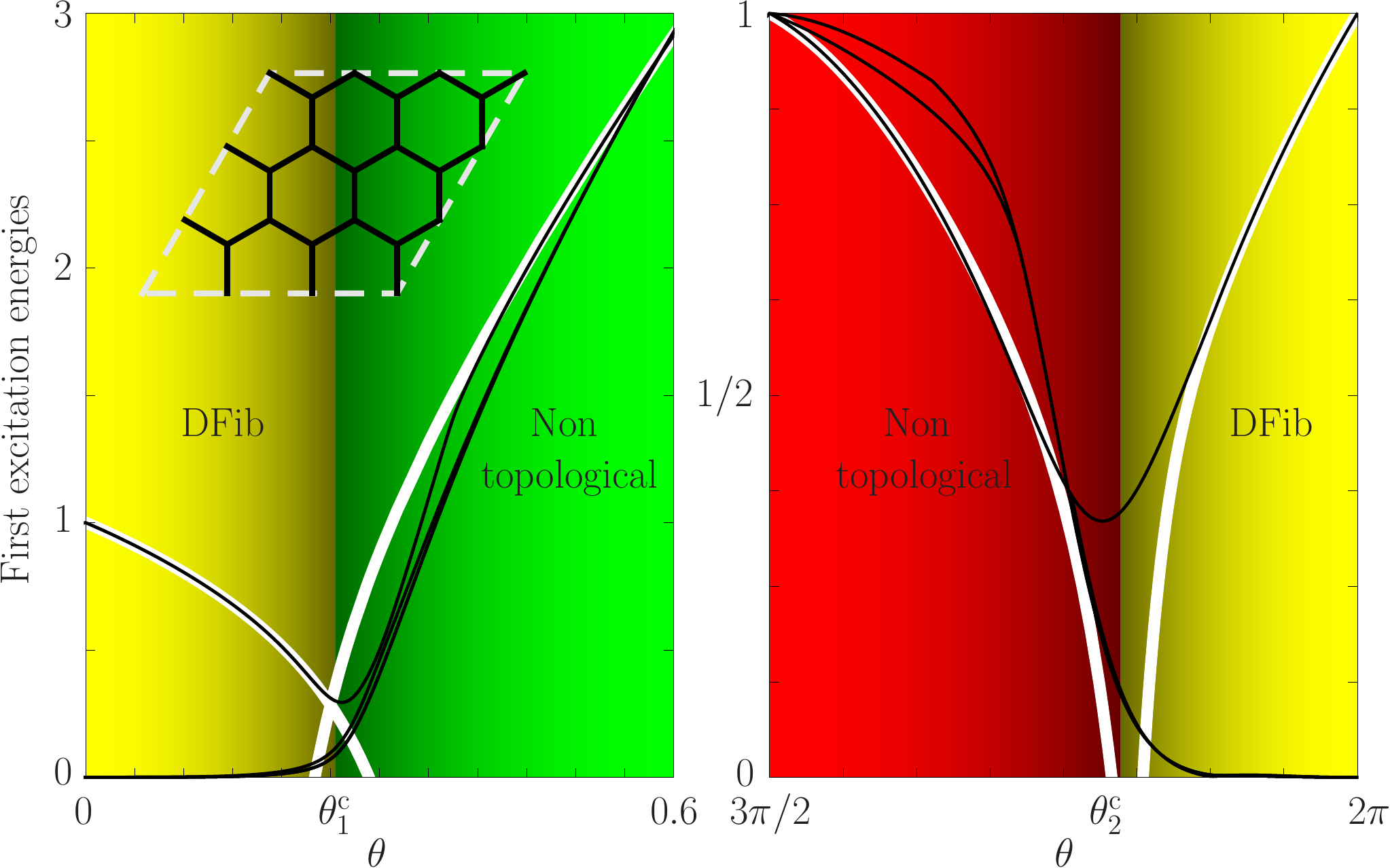}
\caption{(color online). First four excitation energies obtained from the ED results (black lines) for $N_{\rm p}=3\times3$ plaquettes  compared with the low-energy gap computed from high-order bare series expansions  (white lines) given in the Supplemental Material. The topological degeneracy splitting is clearly observed in the vicinity of the critical points. For symmetry reasons, this splitting is only partial for the system considered here (see inset).}
\label{fig:gap}
\end{figure}

%
%
\emph{Phase diagram.---}
%
%
To determine the zero-temperature phase diagram, we combined two
different approaches. First, we performed high-order series expansions in the
thermodynamical limit by means of several methods \cite{Loewdin62,Takahashi77,Knetter00}, around the exactly solvable points $\theta=0$, $\pi/2$,
and $3\pi/2$ described above. This yields the ground-state energy per plaquette
$e_0$ as well as the quasiparticle dispersion from which the low-energy gap
$\Delta$ can be extracted.
Lengthy expressions of these series expansions can be found in the 
Supplemental Material. This method allows one to accurately compute  the critical
couplings for which the gap vanishes. These  points are associated to
second-order transitions but might not be relevant if first-order transitions are
present (see Refs.~\onlinecite{Dusuel11,Schulz12} for details about this issue
in a similar context). Second, we perform
exact diagonalizations (ED) for lattices with periodic boundary conditions. As can be seen in Figs.~\ref{fig:gse} and \ref{fig:gap}, series expansions and ED data are in very good agreement except in the vicinity of the transition points where finite-order and finite-size effects are 
important.

%
%
%
%

\begin{figure}[t]
\includegraphics[width=0.9\columnwidth]{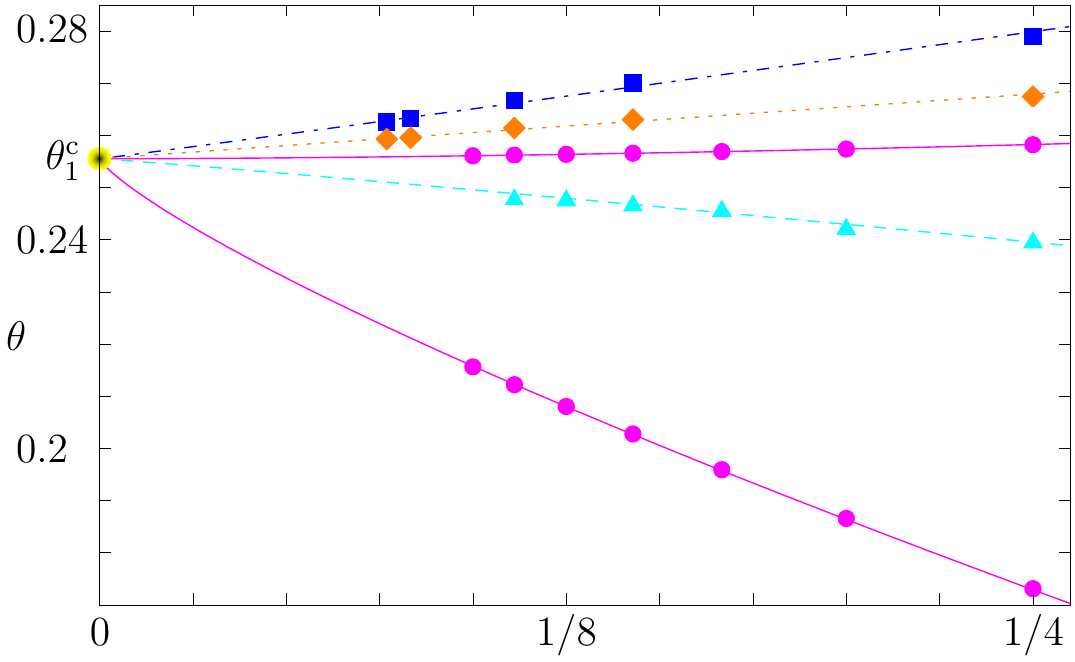}
\caption{(color online). Position of different quantities as a function of either the inverse order $n^{-1}$ of the corresponding series expansion [crossing points of $e_0$ (dots), crossing points of the gap (triangles)] or $N_{\rm p}^{-1}$ [minimum of the gap (squares), minimum of $\partial_\theta^2 e_0$ (diamonds)]. Lines are power-law fits obtained by choosing $\theta^{\rm c}_1$ such that it maximizes the correlations for the crossing points of $e_0$ (solid lines) that are the most accurate results in this work. These crossing points have been obtained from order $n$ series around $\theta=0$ and order $2n-1$ series around $\theta=\pi/2$. For the crossing points of the gap, we used order $n$ series around both limits.}
\label{fig:scaling}
\end{figure}

Combining these two methods (ED and series expansions) we found that {\em the DFib topological phase near $\theta=0$ ranges from 
\mbox{$\theta^{\rm c}_2\simeq -0.63$ $(=5.65)$} to $\theta^{\rm c}_1 \simeq 0.255$.} As we shall now argue, we associate these two points to second-order transitions. 
The first piece of evidence pleading in favor of such a scenario follows from
ED and is the behavior of $\partial_\theta^2 e_0$ that clearly
decreases with the system size near these points (see left and right panels in
Fig.~\ref{fig:dgse}). In addition, the position of the low-energy gap minimum
as well as the topological degeneracy splitting shown in Fig.~\ref{fig:gap} lie
in the same region as the position of the minimum of $\partial_\theta^2 e_0$.
Let us also note that we did not find any relevant level crossing in the
excitation spectrum that could lead to a first-order transition. 
%
%
%
%
%
%
The second argument comes from the high-order perturbation theory.
As can be seen in the Supplemental Material, the series behave very differently for positive and negative $J_{\rm e}$. So, let us first discuss the most favorable case $J_{\rm e}>0$ ($J_{\rm p}>0$).
A close inspection of the series expansion near $\theta=0$ and
$\theta=\pi/2$ reveals three important features: (i)  the sign of each
coefficient is the same in all series; (ii) series of the ground-state
energy intersect in two order-dependent points --- the possible merging of these points, in the infinite-order limit, being a signal of a second-order transition since series then have to be tangential at the critical point;
(iii) series of the gap intersect in a unique (still order-dependent) point, see, e.g., the left part of Fig.~\ref{fig:gap}.
We emphasize that the value of the gap at this crossing point decreases when the order increases and eventually vanishes in the infinite-order limit. In Fig.~\ref{fig:scaling}, we plotted the position of these
crossing points as a function of the (inverse) order as well as the position of
the minimum of the low-energy gap and of $\partial_\theta^2 e_0$ as a function
of $N_{\rm p}^{-1}$ computed from ED results. As can be seen, all data seem to
converge to the same point $\theta^{\rm c}_1\in [0.255,0.256]$ in the infinite-order (size) limit, providing a smoking-gun
evidence of a second-order \mbox{transition}. 

%
%
%
%

Unfortunately, the case $J_{\rm e}<0$ ($J_{\rm p}>0$) is more involved for three
reasons. First, series expansions of $e_0$ and $\Delta$ in this region have alternate signs so that the previous criteria based on crossing points cannot be used. Second, contrary to the case $J_{\rm e}>0$, the momentum minimizing the dispersion of the low-energy quasiparticles is not at the $\Gamma$ point and only belongs to the reciprocal lattice of  $3 \, p\times 3 \,q$ 
systems, $(p,q) \in \mathbb{N}^2$. The only system with such characteristics considered in this study is the $3\times 3$ lattice so that one cannot perform any reliable analysis from ED data near $\theta^{\rm c}_2$.
Third, because of the nature of the low-energy states, maximum orders reachable around 
$\theta=3\pi/2$ are smaller than around $\theta=\pi/2$ (see Supplemental Material).
Nevertheless, using the gap series expansion around $\theta=0$, it is possible to perform  (dlogPad\'e) resummations that lead to a position of the critical point $\theta^{\rm c}_2 \in [5.61,5.62]$. 
Note that the same methods for $J_{\rm e}>0$ would lead to $\theta^{\rm c}_1 \in [0.259,0.261]$. This value differs from the
one proposed above only by a few percent and it overestimates the extension of the topological DFib phase. Consequently, as we have no alternative approach, we roughly estimate that $\theta^{\rm c}_2$ lies in the range $[5.6,5.7]$.

%
%
\emph{Critical exponents.---}
%
%
The obvious question that arises next concerns  universality classes associated to
the transition points. In the absence of a local order parameter, the only meaningful critical exponents for topological phase transitions are those associated to the spectrum. Let us remind that for a second-order transition, the gap vanishes, at large linear system size $L$ and at the critical point, as $\Delta\sim L^{-z}$  where $z$ is the dynamical exponent. In the thermodynamical limit, one further has $\Delta\sim |\theta-\theta^{\rm c}|^{z \nu}$ and $\partial_\theta^2 e_0\sim |\theta-\theta^{\rm c}|^{-\alpha}$. 

As already explained, ED are only useful quantitatively around $\theta^{\rm c}_1$ although restricted to systems of small sizes. Using the gap data, we find $z \simeq 1.2$.
 A finite-size analysis of $e_0$ yields a surprisingly good data collapse
for the $3\times 3$ and $\sqrt{13}\times\sqrt{13}$ systems, with $\theta^{\rm c}_1\simeq 0.255$, $z \simeq 1$, $\nu \simeq 0.4$ and $\alpha\simeq 0.8$ \cite{Schulz13}.
The rather large value of $\alpha$ might be responsible for the quality of the data collapse.
We emphasize that these values are compatible with the previous estimate of $\theta^{\rm c}_1$ as well as with the hyperscaling relation $2-\alpha=\nu(2+z)$.
The above exponents are furthermore in agreement with dlogPad\'e resummations of the series expansion around $\theta=0$ which yield $z \nu \in[0.4,0.44]$.
We discard results from resummations around $\theta=\pi/2$ because these results are not as well converged, since for a given order, clusters needed to compute series around this point are twice as small as those needed around $\theta=0$.
Note that as usual, extracting $\alpha$ from series of $e_0$ does not give any conclusive result.
 Concerning the critical behavior at $\theta^{\rm c}_2$, we  only use dlogPad\'e resummation  around $\theta=0$ and we obtain a gap exponent \mbox{$z \nu \in[0.56,0.58]$}.

%
%
\emph{Outlook.---}
%
%
It is difficult to provide some error bars concerning these values.
To estimate these errors, we performed similar series expansion analysis for the Fibonacci ladder (for which exponents are known exactly \cite{Gils09_1}) and for the 2D $\mathbb{Z}_2$ string-net model (having either Ising or $XY$ transitions depending on the sign of $J_{\rm e}$). The results we obtained \cite{Schulz13} lead us to conclude that critical exponents are to be considered with a precision of about $10\%$.
As a conclusion, we found two different second-order transitions with universality classes that, to the best of our knowledge, are as yet unknown in the context of topological phase transitions. Let us mention that a critical DFib wave function has been proposed \cite{Fendley08} but its relevance for the present problem requires further studies \cite{Schulz13}.

To gain more understanding about these transitions, different approaches could be used, e.g., variational methods or Monte Carlo simulations, although a naive
implementation of the latter should suffer from the sign problem.  It would also be worth studying similar models with a DFib phase \cite{Fendley08,Fendley12} as well as different topological phases. Finally, another important issue concerns transitions between two distinct topological phases \cite{Bais09}. 
Given the ubiquity of Fibonacci anyons in many different physics domain such as topological quantum computation, condensed matter, or atomic physics \cite{Lesanovsky12}, we hope that the present work will stimulate
such investigations.


We thank K. Coester, B. Dou\c{c}ot,  M. Kamfor, and \mbox{J.-B.} Zuber for fruitful  discussions. K.~P. Schmidt acknowledges ESF and EuroHorcs for funding through his EURYI.


\begin{thebibliography}{42}%
\makeatletter
\providecommand \@ifxundefined [1]{%
 \@ifx{#1\undefined}
}%
\providecommand \@ifnum [1]{%
 \ifnum #1\expandafter \@firstoftwo
 \else \expandafter \@secondoftwo
 \fi
}%
\providecommand \@ifx [1]{%
 \ifx #1\expandafter \@firstoftwo
 \else \expandafter \@secondoftwo
 \fi
}%
\providecommand \natexlab [1]{#1}%
\providecommand \enquote  [1]{``#1''}%
\providecommand \bibnamefont  [1]{#1}%
\providecommand \bibfnamefont [1]{#1}%
\providecommand \citenamefont [1]{#1}%
\providecommand \href@noop [0]{\@secondoftwo}%
\providecommand \href [0]{\begingroup \@sanitize@url \@href}%
\providecommand \@href[1]{\@@startlink{#1}\@@href}%
\providecommand \@@href[1]{\endgroup#1\@@endlink}%
\providecommand \@sanitize@url [0]{\catcode `\\12\catcode `\$12\catcode
  `\&12\catcode `\#12\catcode `\^12\catcode `\_12\catcode `\%12\relax}%
\providecommand \@@startlink[1]{}%
\providecommand \@@endlink[0]{}%
\providecommand \url  [0]{\begingroup\@sanitize@url \@url }%
\providecommand \@url [1]{\endgroup\@href {#1}{\urlprefix }}%
\providecommand \urlprefix  [0]{URL }%
\providecommand \Eprint [0]{\href }%
\providecommand \doibase [0]{http://dx.doi.org/}%
\providecommand \selectlanguage [0]{\@gobble}%
\providecommand \bibinfo  [0]{\@secondoftwo}%
\providecommand \bibfield  [0]{\@secondoftwo}%
\providecommand \translation [1]{[#1]}%
\providecommand \BibitemOpen [0]{}%
\providecommand \bibitemStop [0]{}%
\providecommand \bibitemNoStop [0]{.\EOS\space}%
\providecommand \EOS [0]{\spacefactor3000\relax}%
\providecommand \BibitemShut  [1]{\csname bibitem#1\endcsname}%
\let\auto@bib@innerbib\@empty
\bibitem [{\citenamefont {Wen}\ \emph {et~al.}(1989)\citenamefont {Wen},
  \citenamefont {Wilczek},\ and\ \citenamefont {Zee}}]{Wen89_1}%
  \BibitemOpen
  \bibfield  {author} {\bibinfo {author} {\bibfnamefont {X.-G.}\ \bibnamefont
  {Wen}}, \bibinfo {author} {\bibfnamefont {F.}~\bibnamefont {Wilczek}}, \ and\
  \bibinfo {author} {\bibfnamefont {A.}~\bibnamefont {Zee}},\ }\href {\doibase
  10.1103/PhysRevB.39.11413} {\bibfield  {journal} {\bibinfo  {journal} {Phys.
  Rev. B}\ }\textbf {\bibinfo {volume} {39}},\ \bibinfo {pages} {11413}
  (\bibinfo {year} {1989})}\BibitemShut {NoStop}%
\bibitem [{\citenamefont {Wen}(1989)}]{Wen89_2}%
  \BibitemOpen
  \bibfield  {author} {\bibinfo {author} {\bibfnamefont {X.-G.}\ \bibnamefont
  {Wen}},\ }\href {\doibase 10.1103/PhysRevB.40.7387} {\bibfield  {journal}
  {\bibinfo  {journal} {Phys. Rev. B}\ }\textbf {\bibinfo {volume} {40}},\
  \bibinfo {pages} {7387} (\bibinfo {year} {1989})}\BibitemShut {NoStop}%
\bibitem [{\citenamefont {Wen}(1990)}]{Wen90_1}%
  \BibitemOpen
  \bibfield  {author} {\bibinfo {author} {\bibfnamefont {X.-G.}\ \bibnamefont
  {Wen}},\ }\href {\doibase 10.1142/S0217979290000139} {\bibfield  {journal}
  {\bibinfo  {journal} {Int. J. Mod. Phys. B}\ }\textbf {\bibinfo {volume}
  {4}},\ \bibinfo {pages} {239} (\bibinfo {year} {1990})}\BibitemShut {NoStop}%
\bibitem [{\citenamefont {Wen}()}]{Wen12}%
  \BibitemOpen
  \bibfield  {author} {\bibinfo {author} {\bibfnamefont {X.-G.}\ \bibnamefont
  {Wen}},\ }\href@noop {} {}\bibinfo {note}
  {\href{http://arxiv.org/abs/1210.1281}{ arXiv:1210.1281}}\BibitemShut
  {NoStop}%
\bibitem [{\citenamefont {Kitaev}(2003)}]{Kitaev03}%
  \BibitemOpen
  \bibfield  {author} {\bibinfo {author} {\bibfnamefont {A.~Y.}\ \bibnamefont
  {Kitaev}},\ }\href {\doibase 10.1016/S0003-4916(02)00018-0} {\bibfield
  {journal} {\bibinfo  {journal} {Ann. Phys.}\ }\textbf {\bibinfo {volume}
  {303}},\ \bibinfo {pages} {2} (\bibinfo {year} {2003})}\BibitemShut {NoStop}%
\bibitem [{\citenamefont {Bravyi}\ \emph {et~al.}(2010)\citenamefont {Bravyi},
  \citenamefont {Hastings},\ and\ \citenamefont {Michalakis}}]{Bravyi10}%
  \BibitemOpen
  \bibfield  {author} {\bibinfo {author} {\bibfnamefont {S.}~\bibnamefont
  {Bravyi}}, \bibinfo {author} {\bibfnamefont {M.~B.}\ \bibnamefont
  {Hastings}}, \ and\ \bibinfo {author} {\bibfnamefont {S.}~\bibnamefont
  {Michalakis}},\ }\href {\doibase 10.1063/1.3490195} {\bibfield  {journal}
  {\bibinfo  {journal} {J. Math. Phys.}\ }\textbf {\bibinfo {volume} {51}},\
  \bibinfo {pages} {093512} (\bibinfo {year} {2010})}\BibitemShut {NoStop}%
\bibitem [{Pre()}]{Preskill_HP}%
  \BibitemOpen
  \href@noop {} {}\bibinfo {note} {See
  {\href{http://www.theory.caltech.edu/people/preskill/ph219/}{http://www.theory.caltech.edu/people/preskill/ph219/}
  for a pedagogical introduction}}\BibitemShut {NoStop}%
\bibitem [{\citenamefont {Dennis}\ \emph {et~al.}(2002)\citenamefont {Dennis},
  \citenamefont {Kitaev}, \citenamefont {Landahl},\ and\ \citenamefont
  {Preskill}}]{Dennis03}%
  \BibitemOpen
  \bibfield  {author} {\bibinfo {author} {\bibfnamefont {E.}~\bibnamefont
  {Dennis}}, \bibinfo {author} {\bibfnamefont {A.}~\bibnamefont {Kitaev}},
  \bibinfo {author} {\bibfnamefont {A.}~\bibnamefont {Landahl}}, \ and\
  \bibinfo {author} {\bibfnamefont {J.}~\bibnamefont {Preskill}},\ }\href
  {\doibase 10.1063/1.1499754} {\bibfield  {journal} {\bibinfo  {journal} {J.
  Math. Phys.}\ }\textbf {\bibinfo {volume} {43}},\ \bibinfo {pages} {4452}
  (\bibinfo {year} {2002})}\BibitemShut {NoStop}%
\bibitem [{\citenamefont {Dou{\c c}ot}\ and\ \citenamefont
  {Ioffe}(2012)}]{Doucot12}%
  \BibitemOpen
  \bibfield  {author} {\bibinfo {author} {\bibfnamefont {B.}~\bibnamefont
  {Dou{\c c}ot}}\ and\ \bibinfo {author} {\bibfnamefont {L.~B.}\ \bibnamefont
  {Ioffe}},\ }\href {\doibase 10.1088/0034-4885/75/7/072001} {\bibfield
  {journal} {\bibinfo  {journal} {Rep. Prog. Phys.}\ }\textbf {\bibinfo
  {volume} {75}},\ \bibinfo {pages} {072001} (\bibinfo {year}
  {2012})}\BibitemShut {NoStop}%
\bibitem [{\citenamefont {Wen}(2003)}]{Wen03}%
  \BibitemOpen
  \bibfield  {author} {\bibinfo {author} {\bibfnamefont {X.-G.}\ \bibnamefont
  {Wen}},\ }\href {\doibase 10.1103/PhysRevLett.90.016803} {\bibfield
  {journal} {\bibinfo  {journal} {Phys. Rev. Lett.}\ }\textbf {\bibinfo
  {volume} {90}},\ \bibinfo {pages} {016803} (\bibinfo {year}
  {2003})}\BibitemShut {NoStop}%
\bibitem [{\citenamefont {Levin}\ and\ \citenamefont {Wen}(2005)}]{Levin05}%
  \BibitemOpen
  \bibfield  {author} {\bibinfo {author} {\bibfnamefont {M.~A.}\ \bibnamefont
  {Levin}}\ and\ \bibinfo {author} {\bibfnamefont {X.-G.}\ \bibnamefont
  {Wen}},\ }\href {\doibase 10.1103/PhysRevB.71.045110} {\bibfield  {journal}
  {\bibinfo  {journal} {Phys. Rev. B}\ }\textbf {\bibinfo {volume} {71}},\
  \bibinfo {pages} {045110} (\bibinfo {year} {2005})}\BibitemShut {NoStop}%
\bibitem [{\citenamefont {Trebst}\ \emph {et~al.}(2007)\citenamefont {Trebst},
  \citenamefont {Werner}, \citenamefont {Troyer}, \citenamefont {Shtengel},\
  and\ \citenamefont {Nayak}}]{Trebst07}%
  \BibitemOpen
  \bibfield  {author} {\bibinfo {author} {\bibfnamefont {S.}~\bibnamefont
  {Trebst}}, \bibinfo {author} {\bibfnamefont {P.}~\bibnamefont {Werner}},
  \bibinfo {author} {\bibfnamefont {M.}~\bibnamefont {Troyer}}, \bibinfo
  {author} {\bibfnamefont {K.}~\bibnamefont {Shtengel}}, \ and\ \bibinfo
  {author} {\bibfnamefont {C.}~\bibnamefont {Nayak}},\ }\href {\doibase
  10.1103/PhysRevLett.98.070602} {\bibfield  {journal} {\bibinfo  {journal}
  {Phys. Rev. Lett.}\ }\textbf {\bibinfo {volume} {98}},\ \bibinfo {pages}
  {070602} (\bibinfo {year} {2007})}\BibitemShut {NoStop}%
\bibitem [{\citenamefont {Hamma}\ and\ \citenamefont {Lidar}(2008)}]{Hamma08}%
  \BibitemOpen
  \bibfield  {author} {\bibinfo {author} {\bibfnamefont {A.}~\bibnamefont
  {Hamma}}\ and\ \bibinfo {author} {\bibfnamefont {D.~A.}\ \bibnamefont
  {Lidar}},\ }\href {\doibase 10.1103/PhysRevLett.100.030502} {\bibfield
  {journal} {\bibinfo  {journal} {Phys. Rev. Lett.}\ }\textbf {\bibinfo
  {volume} {100}},\ \bibinfo {pages} {030502} (\bibinfo {year}
  {2008})}\BibitemShut {NoStop}%
\bibitem [{\citenamefont {Yu}\ \emph {et~al.}(2008)\citenamefont {Yu},
  \citenamefont {Kou},\ and\ \citenamefont {Wen}}]{Kou08}%
  \BibitemOpen
  \bibfield  {author} {\bibinfo {author} {\bibfnamefont {J.}~\bibnamefont
  {Yu}}, \bibinfo {author} {\bibfnamefont {S.-P.}\ \bibnamefont {Kou}}, \ and\
  \bibinfo {author} {\bibfnamefont {X.-G.}\ \bibnamefont {Wen}},\ }\href
  {\doibase 10.1209/0295-5075/84/17004} {\bibfield  {journal} {\bibinfo
  {journal} {Europhys. Lett.}\ }\textbf {\bibinfo {volume} {84}},\ \bibinfo
  {pages} {17004} (\bibinfo {year} {2008})}\BibitemShut {NoStop}%
\bibitem [{\citenamefont {Vidal}\ \emph
  {et~al.}(2009{\natexlab{a}})\citenamefont {Vidal}, \citenamefont {Dusuel},\
  and\ \citenamefont {Schmidt}}]{Vidal09_1}%
  \BibitemOpen
  \bibfield  {author} {\bibinfo {author} {\bibfnamefont {J.}~\bibnamefont
  {Vidal}}, \bibinfo {author} {\bibfnamefont {S.}~\bibnamefont {Dusuel}}, \
  and\ \bibinfo {author} {\bibfnamefont {K.~P.}\ \bibnamefont {Schmidt}},\
  }\href {\doibase 10.1103/PhysRevB.79.033109} {\bibfield  {journal} {\bibinfo
  {journal} {Phys. Rev. B}\ }\textbf {\bibinfo {volume} {79}},\ \bibinfo
  {pages} {033109} (\bibinfo {year} {2009}{\natexlab{a}})}\BibitemShut
  {NoStop}%
\bibitem [{\citenamefont {Vidal}\ \emph
  {et~al.}(2009{\natexlab{b}})\citenamefont {Vidal}, \citenamefont {Thomale},
  \citenamefont {Schmidt},\ and\ \citenamefont {Dusuel}}]{Vidal09_2}%
  \BibitemOpen
  \bibfield  {author} {\bibinfo {author} {\bibfnamefont {J.}~\bibnamefont
  {Vidal}}, \bibinfo {author} {\bibfnamefont {R.}~\bibnamefont {Thomale}},
  \bibinfo {author} {\bibfnamefont {K.~P.}\ \bibnamefont {Schmidt}}, \ and\
  \bibinfo {author} {\bibfnamefont {S.}~\bibnamefont {Dusuel}},\ }\href
  {\doibase 10.1103/PhysRevB.80.081104} {\bibfield  {journal} {\bibinfo
  {journal} {Phys. Rev. B}\ }\textbf {\bibinfo {volume} {80}},\ \bibinfo
  {pages} {081104} (\bibinfo {year} {2009}{\natexlab{b}})}\BibitemShut
  {NoStop}%
\bibitem [{\citenamefont {Dusuel}\ \emph {et~al.}(2011)\citenamefont {Dusuel},
  \citenamefont {Kamfor}, \citenamefont {Or\'us}, \citenamefont {Schmidt},\
  and\ \citenamefont {Vidal}}]{Dusuel11}%
  \BibitemOpen
  \bibfield  {author} {\bibinfo {author} {\bibfnamefont {S.}~\bibnamefont
  {Dusuel}}, \bibinfo {author} {\bibfnamefont {M.}~\bibnamefont {Kamfor}},
  \bibinfo {author} {\bibfnamefont {R.}~\bibnamefont {Or\'us}}, \bibinfo
  {author} {\bibfnamefont {K.~P.}\ \bibnamefont {Schmidt}}, \ and\ \bibinfo
  {author} {\bibfnamefont {J.}~\bibnamefont {Vidal}},\ }\href {\doibase
  10.1103/PhysRevLett.106.107203} {\bibfield  {journal} {\bibinfo  {journal}
  {Phys. Rev. Lett.}\ }\textbf {\bibinfo {volume} {106}},\ \bibinfo {pages}
  {107203} (\bibinfo {year} {2011})}\BibitemShut {NoStop}%
\bibitem [{\citenamefont {Wu}\ \emph {et~al.}(2012)\citenamefont {Wu},
  \citenamefont {Deng},\ and\ \citenamefont {Prokof'ev}}]{Wu12}%
  \BibitemOpen
  \bibfield  {author} {\bibinfo {author} {\bibfnamefont {F.}~\bibnamefont
  {Wu}}, \bibinfo {author} {\bibfnamefont {Y.}~\bibnamefont {Deng}}, \ and\
  \bibinfo {author} {\bibfnamefont {N.}~\bibnamefont {Prokof'ev}},\ }\href
  {\doibase 10.1103/PhysRevB.85.195104} {\bibfield  {journal} {\bibinfo
  {journal} {Phys. Rev. B}\ }\textbf {\bibinfo {volume} {85}},\ \bibinfo
  {pages} {195104} (\bibinfo {year} {2012})}\BibitemShut {NoStop}%
\bibitem [{\citenamefont {Schulz}\ \emph {et~al.}(2012)\citenamefont {Schulz},
  \citenamefont {Dusuel}, \citenamefont {Or{\'u}s}, \citenamefont {Vidal},\
  and\ \citenamefont {Schmidt}}]{Schulz12}%
  \BibitemOpen
  \bibfield  {author} {\bibinfo {author} {\bibfnamefont {M.~D.}\ \bibnamefont
  {Schulz}}, \bibinfo {author} {\bibfnamefont {S.}~\bibnamefont {Dusuel}},
  \bibinfo {author} {\bibfnamefont {R.}~\bibnamefont {Or{\'u}s}}, \bibinfo
  {author} {\bibfnamefont {J.}~\bibnamefont {Vidal}}, \ and\ \bibinfo {author}
  {\bibfnamefont {K.~P.}\ \bibnamefont {Schmidt}},\ }\href {\doibase
  10.1088/1367-2630/14/2/025005} {\bibfield  {journal} {\bibinfo  {journal}
  {New J. Phys.}\ }\textbf {\bibinfo {volume} {14}},\ \bibinfo {pages} {025005}
  (\bibinfo {year} {2012})}\BibitemShut {NoStop}%
\bibitem [{\citenamefont {Gils}\ \emph {et~al.}(2009)\citenamefont {Gils},
  \citenamefont {Trebst}, \citenamefont {Kitaev}, \citenamefont {Ludwig},
  \citenamefont {Troyer},\ and\ \citenamefont {Wang}}]{Gils09_1}%
  \BibitemOpen
  \bibfield  {author} {\bibinfo {author} {\bibfnamefont {C.}~\bibnamefont
  {Gils}}, \bibinfo {author} {\bibfnamefont {S.}~\bibnamefont {Trebst}},
  \bibinfo {author} {\bibfnamefont {A.}~\bibnamefont {Kitaev}}, \bibinfo
  {author} {\bibfnamefont {A.~W.~W.}\ \bibnamefont {Ludwig}}, \bibinfo {author}
  {\bibfnamefont {M.}~\bibnamefont {Troyer}}, \ and\ \bibinfo {author}
  {\bibfnamefont {Z.}~\bibnamefont {Wang}},\ }\href {\doibase
  10.1038/nphys1396} {\bibfield  {journal} {\bibinfo  {journal} {Nat. Phys.}\
  }\textbf {\bibinfo {volume} {5}},\ \bibinfo {pages} {834} (\bibinfo {year}
  {2009})}\BibitemShut {NoStop}%
\bibitem [{Gil()}]{Gils09_3}%
  \BibitemOpen
  \href@noop {} {}\bibinfo {note} {C. Gils,
  \href{http://iopscience.iop.org/1742-5468/2009/07/P07019/}{J. Stat. Mech.
  P07019 (2009)}}\BibitemShut {NoStop}%
\bibitem [{\citenamefont {Ludwig}\ \emph {et~al.}(2011)\citenamefont {Ludwig},
  \citenamefont {Poilblanc}, \citenamefont {Trebst},\ and\ \citenamefont
  {Troyer}}]{Ludwig11}%
  \BibitemOpen
  \bibfield  {author} {\bibinfo {author} {\bibfnamefont {A.~W.~W.}\
  \bibnamefont {Ludwig}}, \bibinfo {author} {\bibfnamefont {D.}~\bibnamefont
  {Poilblanc}}, \bibinfo {author} {\bibfnamefont {S.}~\bibnamefont {Trebst}}, \
  and\ \bibinfo {author} {\bibfnamefont {M.}~\bibnamefont {Troyer}},\ }\href
  {\doibase 10.1088/1367-2630/13/4/045014} {\bibfield  {journal} {\bibinfo
  {journal} {New J. Phys.}\ }\textbf {\bibinfo {volume} {13}},\ \bibinfo
  {pages} {045014} (\bibinfo {year} {2011})}\BibitemShut {NoStop}%
\bibitem [{\citenamefont {Poilblanc}\ \emph {et~al.}(2011)\citenamefont
  {Poilblanc}, \citenamefont {Ludwig}, \citenamefont {Trebst},\ and\
  \citenamefont {Troyer}}]{Poilblanc11}%
  \BibitemOpen
  \bibfield  {author} {\bibinfo {author} {\bibfnamefont {D.}~\bibnamefont
  {Poilblanc}}, \bibinfo {author} {\bibfnamefont {A.~W.~W.}\ \bibnamefont
  {Ludwig}}, \bibinfo {author} {\bibfnamefont {S.}~\bibnamefont {Trebst}}, \
  and\ \bibinfo {author} {\bibfnamefont {M.}~\bibnamefont {Troyer}},\ }\href
  {\doibase 10.1103/PhysRevB.83.134439} {\bibfield  {journal} {\bibinfo
  {journal} {Phys. Rev. B}\ }\textbf {\bibinfo {volume} {83}},\ \bibinfo
  {pages} {134439} (\bibinfo {year} {2011})}\BibitemShut {NoStop}%
\bibitem [{\citenamefont {Freedman}\ \emph {et~al.}(2012)\citenamefont
  {Freedman}, \citenamefont {Gukelberger}, \citenamefont {Hastings},
  \citenamefont {Trebst}, \citenamefont {Troyer},\ and\ \citenamefont
  {Wang}}]{Freedman12}%
  \BibitemOpen
  \bibfield  {author} {\bibinfo {author} {\bibfnamefont {M.~H.}\ \bibnamefont
  {Freedman}}, \bibinfo {author} {\bibfnamefont {J.}~\bibnamefont
  {Gukelberger}}, \bibinfo {author} {\bibfnamefont {M.~B.}\ \bibnamefont
  {Hastings}}, \bibinfo {author} {\bibfnamefont {S.}~\bibnamefont {Trebst}},
  \bibinfo {author} {\bibfnamefont {M.}~\bibnamefont {Troyer}}, \ and\ \bibinfo
  {author} {\bibfnamefont {Z.}~\bibnamefont {Wang}},\ }\href {\doibase
  10.1103/PhysRevB.85.045414} {\bibfield  {journal} {\bibinfo  {journal} {Phys.
  Rev. B}\ }\textbf {\bibinfo {volume} {85}},\ \bibinfo {pages} {045414}
  (\bibinfo {year} {2012})}\BibitemShut {NoStop}%
\bibitem [{\citenamefont {Burnell}\ \emph {et~al.}(2012)\citenamefont
  {Burnell}, \citenamefont {Simon},\ and\ \citenamefont
  {Slingerland}}]{Burnell12}%
  \BibitemOpen
  \bibfield  {author} {\bibinfo {author} {\bibfnamefont {F.~J.}\ \bibnamefont
  {Burnell}}, \bibinfo {author} {\bibfnamefont {S.~H.}\ \bibnamefont {Simon}},
  \ and\ \bibinfo {author} {\bibfnamefont {J.~K.}\ \bibnamefont
  {Slingerland}},\ }\href {\doibase 10.1088/1367-2630/14/1/015004} {\bibfield
  {journal} {\bibinfo  {journal} {New J. Phys.}\ }\textbf {\bibinfo {volume}
  {14}},\ \bibinfo {pages} {015004} (\bibinfo {year} {2012})}\BibitemShut
  {NoStop}%
\bibitem [{\citenamefont {Simon}\ and\ \citenamefont
  {Fendley}(2013)}]{Simon12}%
  \BibitemOpen
  \bibfield  {author} {\bibinfo {author} {\bibfnamefont {S.~H.}\ \bibnamefont
  {Simon}}\ and\ \bibinfo {author} {\bibfnamefont {P.}~\bibnamefont
  {Fendley}},\ }\href {\doibase 10.1088/1751-8113/46/10/105002} {\bibfield
  {journal} {\bibinfo  {journal} {J. Phys. A}\ }\textbf {\bibinfo {volume}
  {46}},\ \bibinfo {pages} {105002} (\bibinfo {year} {2013})}\BibitemShut
  {NoStop}%
\bibitem [{\citenamefont {Hu}\ \emph {et~al.}(2012)\citenamefont {Hu},
  \citenamefont {Stirling},\ and\ \citenamefont {Wu}}]{Hu12}%
  \BibitemOpen
  \bibfield  {author} {\bibinfo {author} {\bibfnamefont {Y.}~\bibnamefont
  {Hu}}, \bibinfo {author} {\bibfnamefont {S.~D.}\ \bibnamefont {Stirling}}, \
  and\ \bibinfo {author} {\bibfnamefont {Y.-S.}\ \bibnamefont {Wu}},\ }\href
  {\doibase 10.1103/PhysRevB.85.075107} {\bibfield  {journal} {\bibinfo
  {journal} {Phys. Rev. B}\ }\textbf {\bibinfo {volume} {85}},\ \bibinfo
  {pages} {075107} (\bibinfo {year} {2012})}\BibitemShut {NoStop}%
\bibitem [{\citenamefont {{M. D. Schulz {\it et al.}}}()}]{Schulz13}%
  \BibitemOpen
  \bibfield  {author} {\bibinfo {author} {\bibnamefont {{M. D. Schulz {\it et
  al.}}}},\ }\href@noop {} {}\bibinfo {note} {{(to be published).}}\BibitemShut
  {Stop}%
\bibitem [{\citenamefont {Fidkowski}\ \emph {et~al.}(2009)\citenamefont
  {Fidkowski}, \citenamefont {Freedman}, \citenamefont {Nayak}, \citenamefont
  {Walker},\ and\ \citenamefont {Wang}}]{Fidkowski09}%
  \BibitemOpen
  \bibfield  {author} {\bibinfo {author} {\bibfnamefont {L.}~\bibnamefont
  {Fidkowski}}, \bibinfo {author} {\bibfnamefont {M.}~\bibnamefont {Freedman}},
  \bibinfo {author} {\bibfnamefont {C.}~\bibnamefont {Nayak}}, \bibinfo
  {author} {\bibfnamefont {K.}~\bibnamefont {Walker}}, \ and\ \bibinfo {author}
  {\bibfnamefont {Z.}~\bibnamefont {Wang}},\ }\href {\doibase
  10.1007/s00220-009-0757-9} {\bibfield  {journal} {\bibinfo  {journal}
  {Commun. Math. Phys.}\ }\textbf {\bibinfo {volume} {287}},\ \bibinfo {pages}
  {805} (\bibinfo {year} {2009})}\BibitemShut {NoStop}%
\bibitem [{Wan()}]{Wang_book}%
  \BibitemOpen
  \href@noop {} {}\bibinfo {note} {{Z. Wang, {\it Topological Quantum
  Computation}, CBMS Regional Conference Series in Mathematics, Number 112
  (2010)}}\BibitemShut {NoStop}%
\bibitem [{\citenamefont {Burnell}\ and\ \citenamefont
  {Simon}(2010)}]{Burnell10}%
  \BibitemOpen
  \bibfield  {author} {\bibinfo {author} {\bibfnamefont {F.~J.}\ \bibnamefont
  {Burnell}}\ and\ \bibinfo {author} {\bibfnamefont {S.~H.}\ \bibnamefont
  {Simon}},\ }\href {\doibase 10.1016/j.aop.2010.06.003} {\bibfield  {journal}
  {\bibinfo  {journal} {Ann. Phys.}\ }\textbf {\bibinfo {volume} {325}},\
  \bibinfo {pages} {2550} (\bibinfo {year} {2010})}\BibitemShut {NoStop}%
\bibitem [{\citenamefont {Burnell}\ and\ \citenamefont
  {Simon}(2011)}]{Burnell11_1}%
  \BibitemOpen
  \bibfield  {author} {\bibinfo {author} {\bibfnamefont {F.~J.}\ \bibnamefont
  {Burnell}}\ and\ \bibinfo {author} {\bibfnamefont {S.~H.}\ \bibnamefont
  {Simon}},\ }\href {\doibase 10.1088/1367-2630/13/6/065001} {\bibfield
  {journal} {\bibinfo  {journal} {New J. Phys.}\ }\textbf {\bibinfo {volume}
  {13}},\ \bibinfo {pages} {065001} (\bibinfo {year} {2011})}\BibitemShut
  {NoStop}%
\bibitem [{\citenamefont {Burnell}\ \emph {et~al.}(2011)\citenamefont
  {Burnell}, \citenamefont {Simon},\ and\ \citenamefont
  {Slingerland}}]{Burnell11_2}%
  \BibitemOpen
  \bibfield  {author} {\bibinfo {author} {\bibfnamefont {F.~J.}\ \bibnamefont
  {Burnell}}, \bibinfo {author} {\bibfnamefont {S.~H.}\ \bibnamefont {Simon}},
  \ and\ \bibinfo {author} {\bibfnamefont {J.~K.}\ \bibnamefont
  {Slingerland}},\ }\href {\doibase 10.1103/PhysRevB.84.125434} {\bibfield
  {journal} {\bibinfo  {journal} {Phys. Rev. B}\ }\textbf {\bibinfo {volume}
  {84}},\ \bibinfo {pages} {125434} (\bibinfo {year} {2011})}\BibitemShut
  {NoStop}%
\bibitem [{\citenamefont {Koenig}\ \emph {et~al.}(2010)\citenamefont {Koenig},
  \citenamefont {Kuperberg},\ and\ \citenamefont {Reichardt}}]{Koenig10}%
  \BibitemOpen
  \bibfield  {author} {\bibinfo {author} {\bibfnamefont {R.}~\bibnamefont
  {Koenig}}, \bibinfo {author} {\bibfnamefont {G.}~\bibnamefont {Kuperberg}}, \
  and\ \bibinfo {author} {\bibfnamefont {B.~W.}\ \bibnamefont {Reichardt}},\
  }\href {\doibase 10.1016/j.aop.2010.08.001} {\bibfield  {journal} {\bibinfo
  {journal} {Ann. Phys.}\ }\textbf {\bibinfo {volume} {325}},\ \bibinfo {pages}
  {2707} (\bibinfo {year} {2010})}\BibitemShut {NoStop}%
\bibitem [{\citenamefont {Bais}\ and\ \citenamefont
  {Slingerland}(2009)}]{Bais09}%
  \BibitemOpen
  \bibfield  {author} {\bibinfo {author} {\bibfnamefont {F.~A.}\ \bibnamefont
  {Bais}}\ and\ \bibinfo {author} {\bibfnamefont {J.~K.}\ \bibnamefont
  {Slingerland}},\ }\href {\doibase 10.1103/PhysRevB.79.045316} {\bibfield
  {journal} {\bibinfo  {journal} {Phys. Rev. B}\ }\textbf {\bibinfo {volume}
  {79}},\ \bibinfo {pages} {045316} (\bibinfo {year} {2009})}\BibitemShut
  {NoStop}%
\bibitem [{\citenamefont {Nussinov}\ and\ \citenamefont
  {Ortiz}(2009)}]{Nussinov09}%
  \BibitemOpen
  \bibfield  {author} {\bibinfo {author} {\bibfnamefont {Z.}~\bibnamefont
  {Nussinov}}\ and\ \bibinfo {author} {\bibfnamefont {G.}~\bibnamefont
  {Ortiz}},\ }\href {\doibase 10.1016/j.aop.2008.11.002} {\bibfield  {journal}
  {\bibinfo  {journal} {Ann. Phys.}\ }\textbf {\bibinfo {volume} {324}},\
  \bibinfo {pages} {977} (\bibinfo {year} {2009})}\BibitemShut {NoStop}%
\bibitem [{\citenamefont {L\"{o}wdin}(1962)}]{Loewdin62}%
  \BibitemOpen
  \bibfield  {author} {\bibinfo {author} {\bibfnamefont {P.-O.}\ \bibnamefont
  {L\"{o}wdin}},\ }\href {\doibase 10.1063/1.1724312} {\bibfield  {journal}
  {\bibinfo  {journal} {J. Math. Phys.}\ }\textbf {\bibinfo {volume} {3}},\
  \bibinfo {pages} {969} (\bibinfo {year} {1962})}\BibitemShut {NoStop}%
\bibitem [{\citenamefont {Takahashi}(1977)}]{Takahashi77}%
  \BibitemOpen
  \bibfield  {author} {\bibinfo {author} {\bibfnamefont {M.}~\bibnamefont
  {Takahashi}},\ }\href {\doibase doi:10.1088/0022-3719/10/8/031} {\bibfield
  {journal} {\bibinfo  {journal} {J. Phys. C}\ }\textbf {\bibinfo {volume}
  {10}},\ \bibinfo {pages} {1289} (\bibinfo {year} {1977})}\BibitemShut
  {NoStop}%
\bibitem [{\citenamefont {Knetter}\ and\ \citenamefont
  {Uhrig}(2000)}]{Knetter00}%
  \BibitemOpen
  \bibfield  {author} {\bibinfo {author} {\bibfnamefont {C.}~\bibnamefont
  {Knetter}}\ and\ \bibinfo {author} {\bibfnamefont {G.~S.}\ \bibnamefont
  {Uhrig}},\ }\href {\doibase 10.1007/s100510050026} {\bibfield  {journal}
  {\bibinfo  {journal} {Eur. Phys. J. B}\ }\textbf {\bibinfo {volume} {13}},\
  \bibinfo {pages} {209} (\bibinfo {year} {2000})}\BibitemShut {NoStop}%
\bibitem [{\citenamefont {Fendley}(2008)}]{Fendley08}%
  \BibitemOpen
  \bibfield  {author} {\bibinfo {author} {\bibfnamefont {P.}~\bibnamefont
  {Fendley}},\ }\href {\doibase 10.1016/j.aop.2008.04.011} {\bibfield
  {journal} {\bibinfo  {journal} {Ann. Phys.}\ }\textbf {\bibinfo {volume}
  {323}},\ \bibinfo {pages} {3113} (\bibinfo {year} {2008})}\BibitemShut
  {NoStop}%
\bibitem [{\citenamefont {Fendley}\ \emph {et~al.}()\citenamefont {Fendley},
  \citenamefont {Isakov},\ and\ \citenamefont {Troyer}}]{Fendley12}%
  \BibitemOpen
  \bibfield  {author} {\bibinfo {author} {\bibfnamefont {P.}~\bibnamefont
  {Fendley}}, \bibinfo {author} {\bibfnamefont {S.~V.}\ \bibnamefont {Isakov}},
  \ and\ \bibinfo {author} {\bibfnamefont {M.}~\bibnamefont {Troyer}},\
  }\href@noop {} {}\bibinfo {note} {\href{http://arxiv.org/abs/1210.5527}{
  arXiv:1210.5527}}\BibitemShut {NoStop}%
\bibitem [{\citenamefont {Lesanovsky}\ and\ \citenamefont
  {Katsura}(2012)}]{Lesanovsky12}%
  \BibitemOpen
  \bibfield  {author} {\bibinfo {author} {\bibfnamefont {I.}~\bibnamefont
  {Lesanovsky}}\ and\ \bibinfo {author} {\bibfnamefont {H.}~\bibnamefont
  {Katsura}},\ }\href {\doibase 10.1103/PhysRevA.86.041601} {\bibfield
  {journal} {\bibinfo  {journal} {Phys. Rev. A}\ }\textbf {\bibinfo {volume}
  {86}},\ \bibinfo {pages} {041601} (\bibinfo {year} {2012})}\BibitemShut
  {NoStop}%
\end{thebibliography}
%


\onecolumngrid
\newpage

\section*{Supplemental Material}
In the following, we give the series expansions in the different phases for the ground-state energy per plaquette $e_0$ and the quasiparticle gaps $\Delta^{\pm}$, for positive and negative signs of the dimensionless parameter $t=J_{\rm e}/J_{\rm p}=\tan \theta$ respectively. For the sake of clarity, we give below the numerical values of the coefficients with 16 digits.

\subsection{Expansions in the vicinity \texorpdfstring{$\theta=0$}{theta=0}}

The ground-state energy per plaquette $e_0$ near $\theta=0$ ($J_{\rm{p}}=1$, $J_{\rm{e}}=0$) has been obtained up to order $11$ using operator perturbation theory \cite{Takahashi77}, whereas quasiparticle gaps $\Delta^\pm$ were obtained up to order $9$ using perturbative continuous unitary transformations. \cite{Knetter00}

%
%
\begin{align}
	e_0 / J_{\rm{p}} =& \ -1-0.8291796067500631 \,t-0.3 \,t^2-0.2329179606750063 \,t^3-0.3758359213500126 \,t^4\nonumber\\&\
	-0.6934622369921362 \,t^5-1.517757831138397 \,t^6-3.615896887905089 \,t^7-9.257482947753094 \,t^8\nonumber\\&\
	-24.89646210135949 \,t^9-69.63655938933877 \,t^{10}-200.8253697230269 \,t^{11}, \displaybreak[0]\\
	\nonumber\\
 \Delta^+ / J_{\rm{p}} =& \ 1  -1.658359213500126 \,t-2.029179606750063 \,t^2-3.107113095525145 \,t^3-8.042597266963313 \,t^4\nonumber\\&\
 -19.16249885423558 \,t^5  -58.31720409052607 \,t^6-164.4421257647495 \,t^7-528.5318111014412 \,t^8\nonumber\\&\
 -1615.453328025113 \,t^9, \displaybreak[0]\\
 \nonumber\\
 \Delta^- / J_{\rm{p}} =& \ 1 +0.8291796067500631 \,t+0.1145898033750315 \,t^2+0.5110332556124590 \,t^3+0.4044760408194119 \,t^4\nonumber\\&\
 +0.9554382719956335 \,t^5 +1.784752477740017 \,t^6+4.523961920423115 \,t^7+11.17294663306187 \,t^8\nonumber\\&\
 +31.20020681009114 \,t^9.
\end{align}
%
%
\subsection{Expansions in the vicinity \texorpdfstring{$\theta=\pi/2$}{theta=pi/2}}
The ground-state energy per plaquette $e_0$ near $\theta=\pi/2$ ($J_{\rm{e}}=1$, $J_{\rm{p}}=0$) has been obtained up to order $19$ using a partitioning technique provided by L\"owdin \cite{Loewdin62},  whereas quasiparticle gaps $\Delta^\pm$ were obtained up to order $11$ using operator perturbation theory \cite{Takahashi77} on appropriate periodic clusters.
%
%
\begin{align}
e_0 / J_{\rm{e}} =&\ -3-2.763932022500210\cdot 10^{-1} \,t^{-1}-3.333333333333333\cdot 10^{-2} \,t^{-2}-2.484519974999766\cdot 10^{-3} \,t^{-3}\nonumber\\&\
-1.473090114646592\cdot 10^{-4} \,t^{-4}-1.762516450320833\cdot 10^{-5} \,t^{-5}-3.114829150546602\cdot 10^{-6} \,t^{-6}\nonumber\\&\
-4.974954832303385\cdot 10^{-7} \,t^{-7}-8.712942025753695\cdot 10^{-8} \,t^{-8}-1.680470831303724\cdot 10^{-8} \,t^{-9}\nonumber\\&\
-3.252798060742993\cdot 10^{-9} \,t^{-10}-6.452247707667803\cdot 10^{-10} \,t^{-11}-1.328942917494399\cdot 10^{-10} \,t^{-12}\nonumber\\&\
-2.786531244027440\cdot 10^{-11} \,t^{-13}-5.923931315552463\cdot 10^{-12} \,t^{-14}-1.280290681522505\cdot 10^{-12} \,t^{-15}\nonumber\\&\
-2.803432623420322\cdot 10^{-13} \,t^{-16}-6.202995686227467\cdot 10^{-14} \,t^{-17}-1.386248233596245\cdot 10^{-14} \,t^{-18}\nonumber\\&\
-3.125540962122164\cdot 10^{-15} \,t^{-19},\displaybreak[0]\\
\nonumber\\
\Delta^{\pm} / J_{\rm{e}}=&\ 6-4.472135954999579\cdot 10^{-1} \,t^{-1}-5.916925468334595\cdot 10^{-2} \,t^{-2}-1.532368203449658\cdot 10^{-2} \,t^{-3}\nonumber\\&\
-1.826128841179490\cdot 10^{-3} \,t^{-4}-2.141218566177655\cdot 10^{-4} \,t^{-5}-7.057273308076495\cdot 10^{-5} \,t^{-6}\nonumber\\&\
-7.713120260517609\cdot 10^{-6} \,t^{-7}-1.822619200222926\cdot 10^{-6} \,t^{-8}-1.739394034185955\cdot 10^{-7} \,t^{-9}\nonumber\\&\
-1.068512192171555\cdot 10^{-7} \,t^{-10}-1.583732801424164\cdot 10^{-8} \,t^{-11}.
\end{align}
%
%

\subsection{Expansions in the vicinity \texorpdfstring{$\theta=3\pi/2$}{theta=3pi/2}}
The ground-state energy per plaquette $e_0$ near $\theta=3\pi/2$ ($J_{\rm{e}}=-1$, $J_{\rm{p}}=0$) has been obtained up to order $9$ using partitioning techniques \cite{Loewdin62}, whereas  quasiparticle gaps $\Delta^\pm$ were obtained up to order $6$ using perturbative continuous unitary transformations. \cite{Knetter00}
%
%
\begin{align}
\!\!e_0 / \left(-J_{\rm{e}}\right)=&\ 
3.013155617496425\cdot 10^{-1}\,t^{-1}-1.132044933254820\cdot 10^{-1}\,t^{-2}+2.807797460719963\cdot 10^{-2}\,t^{-3}\nonumber\\&\
-4.507809490972001\cdot 10^{-3}\,t^{-4}-3.037988446794379\cdot 10^{-3}\,t^{-5}+4.596752532754356\cdot 10^{-3}\,t^{-6}\nonumber\\&\
-1.633669374413878\cdot 10^{-3}\,t^{-7}-2.288717115668441\cdot 10^{-3}\,t^{-8}+3.812642130493073\cdot 10^{-3}\,t^{-9},\displaybreak[0]\\
\nonumber \\
\!\!\Delta^+ / \left(-J_{\rm{e}}\right) =&\ 
1-7.331262919989905\cdot 10^{-1}\,t^{-1}+2.627515502196252\cdot 10^{-1}\,t^{-2}-1.016556730679853\cdot 10^{-1}\,t^{-3}\nonumber\\&\
+1.966652521926989\cdot 10^{-2}\,t^{-4}-7.070423645089451\cdot 10^{-2}\,t^{-5}+1.098516898950685\cdot 10^{-1}\,t^{-6}, \displaybreak[0]\\
\nonumber \\
\!\!\Delta^- / \left(-J_{\rm{e}}\right) =&\ 
1+2.668737080010095\cdot 10^{-1}\,t^{-1}-2.048165955279984\cdot 10^{-1}\,t^{-2}-7.571315893653220\cdot 10^{-2}\,t^{-3}\nonumber\\&\
+3.964262957878962\cdot 10^{-2}\,t^{-4}+7.257956045336907\cdot 10^{-2}\,t^{-5}-6.691149658050659\cdot 10^{-2}\,t^{-6}.
\end{align}
%
%

\end{document}